\documentclass[twocolumn]{aastex61}

\received{2018}
\revised{2018}
\accepted{July 10, 2018}
\submitjournal{ApJ}

\shorttitle{Near-Infrared Imaging Polarimetry of FUors}
\shortauthors{Takami et al.}

\usepackage{color}
\usepackage{soul,xcolor}
\setstcolor{red}

\begin{document}

\title{Near-Infrared High-Resolution Imaging Polarimetry of FU Ori-Type Objects: Towards A Unified Scheme for Low-Mass Protostellar Evolution\footnote{Based on data collected at Subaru Telescope, which is operated by the National Astronomical Observatory of Japan.}}

\begin{abstract}
We present near-IR imaging polarimetry of five classical FU Ori-type objects (FU Ori, V1057 Cyg, V1515 Cyg, V1735 Cyg, Z CMa)  with a $\sim$0\farcs1 resolution observed using HiCIAO+AO188 at Subaru Telescope. We observed scattered light associated with circumstellar dust around four of them (i.e., all but V1515 Cyg). Their polarized intensity distribution shows a variety of morphologies with arms, tails or streams, spikes and fragmented distributions, many of which were reported in our previous paper. The morphologies of these reflection nebulae significantly differ from many other normal young stellar objects (Class I-II objects). These structures are attributed to gravitationally unstable disks, trails of clump ejections, dust blown by a wind or a jet, and  a stellar companion. We can consistently explain our results with the scenario that their accretion outbursts (FUor outbursts) are triggered by gravitationally fragmenting disks, and with the hypothesis that many low-mass young stellar objects experience such outbursts. 
\end{abstract}

\keywords{accretion, accretion disks --- instabilities --- polarization --- scattering --- stars: formation --- stars: protostars --- stars: winds, outflows}

\correspondingauthor{Michihiro Takami}
\email{hiro@asiaa.sinica.edu,.tw}

\author{Michihiro Takami}
\affil{Institute of Astronomy and Astrophysics, Academia Sinica, 
11F of Astronomy-Mathematics Building, AS/NTU
No.1, Sec. 4, Roosevelt Rd, Taipei 10617, Taiwan, R.O.C.}

\author{Guangwei Fu}
\affil{Institute of Astronomy and Astrophysics, Academia Sinica, 
11F of Astronomy-Mathematics Building, AS/NTU
No.1, Sec. 4, Roosevelt Rd, Taipei 10617, Taiwan, R.O.C.}
\affil{Department of Astronomy, University of Wisconsin - Madison, 2535 Sterling Hall, 475 N. Charter Street, Madison, WI 53706-1507}
\affil{Department of Astronomy, University of Maryland, College Park, MD 20742-2421, USA}

\author{Hauyu Baobab Liu}
\affil{European Southern Observatory (ESO), Karl-Schwarzschild-Strasse 2, D-85748 Garching, Germany}

\author{Jennifer L. Karr}
\affil{Institute of Astronomy and Astrophysics, Academia Sinica, 
11F of Astronomy-Mathematics Building, AS/NTU
No.1, Sec. 4, Roosevelt Rd, Taipei 10617, Taiwan, R.O.C.}

\author{Jun Hashimoto}
\affil{Astrobiology Center of NINS, 2-21-1, Osawa, Mitaka, Tokyo, 181-8588, Japan}

\author{Tomoyuki Kudo}
\affil{Subaru Telescope, National Astronomical Observatory
of Japan, National Institutes of Natural Sciences (NINS), 650 North
A`oh\=ok\=u Place, Hilo, HI 96720, USA}

\author{Eduard I. Vorobyov}
\affil{Department of Astrophysics, The University of Vienna, Vienna, A-1180, Austria}
\affil{Research Institute of Physics, Southern Federal University, Stachki 194, Rostov-on-Don, 344090, Russia}

\author{\'Agnes K\'osp\'al}
\affil{Konkoly Observatory, Research Centre for Astronomy and Earth Sciences, Hungarian Academy of Sciences, Konkoly-Thege Mikl\'os \'ut 15-17, 1121 Budapest, Hungary}
\affil{Max-Planck-Institute for Astronomy, Koenigsstuhl 17, D-69117 Heidelberg, Germany}

\author{Peter Scicluna}
\affil{Institute of Astronomy and Astrophysics, Academia Sinica, 
11F of Astronomy-Mathematics Building, AS/NTU
No.1, Sec. 4, Roosevelt Rd, Taipei 10617, Taiwan, R.O.C.}

\author{Ruobing Dong}
\affil{Steward Observatory, University of Arizona, Tucson, Arizona 85721, USA }

\author{Motohide Tamura}
\affil{Astrobiology Center of NINS, 2-21-1, Osawa, Mitaka, Tokyo, 181-8588, Japan}
\affil{Department of Astronomy, The University of Tokyo, 7-3-1 Hongo, Bunkyo-ku, Tokyo 113-0033, Japan}
\affil{National Astronomical Observatory of Japan, 2-21-1 Osawa, Mitaka, Tokyo 181-8588, Japan}

\author{Tae-Soo Pyo}
\affil{Subaru Telescope, National Astronomical Observatory
of Japan, National Institutes of Natural Sciences (NINS), 650 North
A`oh\=ok\=u Place, Hilo, HI 96720, USA}
\affil{School of Mathematical and Physical Science, The
Graduate University for Advanced Studies (SOKENDAI), Hayama, Kanagawa
240-0193, Japan}

\author{Misato Fukagawa}
\affil{Division of Particle and Astrophysical Science, Graduate School of Science, Nagoya University, Furo-cho, Chikusa-ku, Nagoya, Aich 464-8602, Japan}

\author{Toru Tsuribe}
\affil{College of Science, Ibaraki University, 2-1-1 Bunkyo, Mito, Ibaraki 310-8512, Japan}

\author{Michael M. Dunham}
\affil{Department of Physics, State University of New York at Fredonia, Fredonia, NY 14063, USA}

\author{Thomas Henning}
\affil{Max-Planck-Institute for Astronomy, Koenigsstuhl 17, D-69117 Heidelberg, Germany}

\author{Jerome de Leon}
\affil{Department of Astronomy, The University of Tokyo, 7-3-1 Hongo, Bunkyo-ku, Tokyo 113-0033, Japan}


\section{Introduction}

Most of the stars in our Galaxy have masses below a few solar masses. We do not understand well the physical mechanism by which these stars (``low-mass" stars) accrete their masses. \citet{Muzerolle98_IR} measured the mass accretion rates for a sample of low-mass young stellar objects (Class I-II YSOs) using the Br $\gamma$ line, and showed that steady mass accretion can explain only a fraction of their final stellar masses.  This issue is also corroborated by the facts below. At the pre-main sequence phase of their evolution (namely ``Class II-III''), in which the circumstellar gas+dust envelope has already been dissipated, the masses of the associated circumstellar disks are significantly smaller than the stars \citep[e.g.,][]{Williams11}, indicating that the stellar masses have been developed primarily at younger evolutionary stages: i.e., the ``Class 0-I'' phases. 
It has been suggested that their protostellar luminosities
tend to be significantly lower than theoretical predictions for steady mass accretion, e.g., by a factor of 10-10$^3$ \citep[the ``luminosity problem''; see][for recent reviews]{Dunham14,Audard14}.


A key phenomenon that may solve the above issues is episodic mass accretion, observed in some YSOs as a sudden increase of flux at the optical and near-IR wavelengths. The above trends are explained if Class 0-I YSOs are associated with accretion outbursts whose periods are significantly shorter than the time scale of these evolutionary phases (and therefore with a small chance of observation) but which are responsible for a significant fraction of the final stellar masses \citep[e.g.,][]{Kenyon90,Muzerolle98_IR,Calvet00}. 

The FU Orionis objects (hereafter FUors) are a class of YSOs which undergo the most active and violent accretion outbursts during which the accretion rate rapidly increases by a factor of $\sim$1000, and remains high for several decades or more. Such outbursts have been observed toward about 10 stars to date. Another dozen YSOs exhibit optical or near-IR spectra similar to FUors, distinct from many other YSOs, but outbursts have never been observed. Their spectra suggest disk accretion with high accretion rates similar to FUors. These are classified as FUor candidates or FUor-like objects. See \citet{Audard14} for a recent review for FUors and FUor-like objects. Their optical and near-IR spectra indicate that the optical and near-IR emission from these objects is dominated by a warm disk photosphere \citep[][for reviews]{Hartmann96,Audard14}. This is in contrast to many other normal Class I-II YSOs, whose optical and near-IR emission is dominated by the star and featureless dust continuum \citep[e.g.,][]{Greene96,Doppmann05,Connelley10}. Many FUors are known to host a massive circumstellar gas+dust envelope comparable to normal Class I YSOs \citep[e.g.,][]{Sandell01,Millan-Gabet06,Kospal17b,Kospal17c}, consistent with the above hypothesis that many Class I YSOs experience such accretion outbursts.

The triggering mechanism of the outbursts is not clear, despite numerous observations at a variety of wavelengths, theoretical work, and numerical simulations. The proposed mechanisms include: (1) gravitational/thermal/magneto-rotational instabilities in the disk or (2) the perturbation of the disk by an external body \citep[see][for a recent review]{Audard14}. A sudden increase in the accretion rate heats up the inner disk ($r$$<$1 AU), observable as continuum emission at optical and infrared wavelengths. An infalling envelope or gravitational instability in the outer disk ($r$$\gg$1 AU) may also affect the outbursts \citep[e.g.,][]{Hartmann96,Vorobyov05,Vorobyov10,Vorobyov15b}. 

Near-IR imaging polarimetry is a powerful tool for observing the geometry of circumstellar disks and envelopes associated with YSOs at high angular resolutions. These disks and envelopes contain a number of dust grains which scatters the light from the central source. While the scattered light is significantly fainter than the central source, its large polarization relative to the central object allows us to observe circumstellar structures very close to the central source ($r=$0\farcs1-0\farcs2).

Using this technique, \citet{Liu16} (hereafter Paper I) have recently revealed complicated circumstellar structures associated with classical FUors. These structures include arms similar to those of spirals, elongated structures which may be associated with gas streams or jets, and spiky structures. In Paper I we performed comparisons between these structures and simulations, and showed that some of them agree with those expected for gravitational instabilities in disks, i.e., one of the mechanisms extensively studied to explain FUor outbursts in young low-mass stars and FUor-type outbursts in young massive stars \citep[e.g.,][]{Vorobyov05,Vorobyov10,Vorobyov15b,Machida11,Zhu12_GI,Meyer17,Zhao18,Kueffmeier18}. Gravitational fragmentation in disks may also be the key mechanism for the formation of planets and brown dwarfs \citep[e.g.,][]{Boss03,Nayaksin10,Vorobyov13,Stamatellos15}. 

Near-IR imaging polarimetry is also powerful for investigating the vertical structures of cicrumstellar disks and envelopes. \citet{Kospal08} used this technique for an FUor-like object and revealed the presence of a flat Keplerian disk system embedded in the envelope. In contrast to disks with a face-on view and intermediate inclinations, this edge-on disk system is associated with low polarization due to multiple scattering, agreeing with model predictions \citep[e.g.,][]{Fischer96,Murakawa10}.
		
	In this paper we extend our analysis of the near-IR data partially used for Paper I. The rest of the paper is organized as follows. In Section 2 we describe our observations of near-IR imaging polarimetry with coronagraphy. In Section 3 we show the results. In Section 4 we discuss the origins and implications of the observed structures. In Section 5 we discuss the consistency of our results with the scenario that their accretion outbursts (FUor outbursts) are triggered by gravitational fragmenting disks, and with the hypothesis that many normal low-mass YSOs experience such outbursts. We give a summary in Section 6.
	

\section{Observations and Data Reduction}
	Observations of five FUors were performed using HiCIAO at Subaru Telescope in polarization differential imaging (PDI) mode coupled with adaptive optics (AO188). Tables \ref{tbl:targets} and \ref{tbl:observations} summarize the targets and the observations, respectively.
		A Wollaston prism splits the incident light into two orthogonal linearly polarized beams on the detector. By rotating the half-wave plate to $0^{\circ}, 22.5^{\circ}, 45^{\circ},$ and $67.5^{\circ}$, the linear polarization was measured. 
	The AO guiding stars were the target objects themselves, and the angular resolution (FWHM) was 0\farcs08 on average. A coronagraph with a 0\farcs3 diameter occulting mask was used to block out light from the bright central point source.
In addition to the science frames, each target was observed with short exposures (1.5 s), a single retarder angle ($0^{\circ}$) and without a coronagraphic mask to measure the flux from the central point source.
    

\begin{table*}

\caption{Targets \label{tbl:targets}}
\begin{tabular}{lcccccccr}
\tableline\tableline
Object 	& Distance\tablenotemark{a} & $L_{bol}$\tablenotemark{b}	&  $A_V$\tablenotemark{c}	& Onset of\\
		& (pc)				& ($L_\odot$)	& (mag.) & Burst(s)\tablenotemark{c} \\
\tableline
FU Ori 		& 420	& 100-430\tablenotemark{d,e,f,g}	& 1.5-2.6 	& 1936 \\
V1057 Cyg 	& 900	& 200-500\tablenotemark{d,e,f}		& 3.0-4.2	& 1970 \\
V1735 Cyg  	& 600	& 120\tablenotemark{e}			& 8.0-10.8	& $>$1957,$<$1965 \\
Z CMa 		& 1000	& 400-600\tablenotemark{h}		& 1.8-3.5	& Many \\
V1515 Cyg	& 1000	& 200\tablenotemark{e,f}		& 2.8-3.2	& $\sim$1950 \\
\tableline \tableline
\end{tabular}
\tablenotetext{a}{Based on Gaia Data Release 2 for FU Ori, V1057 Cyg, V1735 Cyg and V1515 Cyg; \citet{Kaltcheva00} for Z CMa. 
The actual values are $416_{-8}^{+9}$, $920_{-31}^{+35}$, $624_{-37}^{+41}$, $990\pm50$, and $1009_{-28}^{+30}$ pc for FU Ori, V1057 Cyg, V1735 Cyg, Z CMa, and V1515 Cyg, respectively. The distance to Z CMa is controvertial (Appendix \ref{app:zcma_distance}).
We will use the values tabulated above for the analysis in the rest of the paper, while we will consider these uncertainties of the distance measurements in Section 3.5.
}
\tablenotetext{b}{Scaled to the distances adopted for this work}
\tablenotetext{c}{\citet{Audard14}}
\tablenotetext{d}{\citet{Harvey82}}
\tablenotetext{e}{\citet{Levreault88}}
\tablenotetext{f}{\citet{Green06}}
\tablenotetext{g}{\citet{Smith82,Adams87}}
\tablenotetext{h}{\citet{Thiebaut95}}
\end{table*}


\begin{table*}
\caption{Observations \label{tbl:observations}}
\begin{tabular}{lccccr}
\tableline\tableline
Object 	& Band & Date  & FOV	& \multicolumn{2}{c}{Polarization of the central source} \\
		&	& (UT)	&&	Degree (\%)	& P.A. (deg.) \\
\tableline
FU Ori 		& $H$	& Oct 5, 2014 	& 20\arcsec$\times$10\arcsec	& 0.24$\pm$0.05	& --42$\pm$5\\
V1057 Cyg 	& $H$	& Oct 5, 2014 	& 20\arcsec$\times$10\arcsec	& 0.7$\pm$0.1		& 68$\pm$2\\
V1735 Cyg 	& $H$	& Oct 6, 2014 	& 20\arcsec$\times$10\arcsec	&1.7	$\pm$0.1		& 18$\pm$1\\
			& $K$	& Oct 6, 2014 	& 20\arcsec$\times$10\arcsec	& 1.0$\pm$0.1		& 18$\pm$5\\
Z CMa 		& $J$	& Oct 6, 2014 	& 5\arcsec$\times$5 \arcsec	& 1.16$\pm$0.05	& --21$\pm$1\\
	 		& $H$	& Oct 6, 2014 	& 5\arcsec$\times$5 \arcsec	& 0.70$\pm$0.05	& --23$\pm$2\\
			& $K$	& Oct 6, 2014 	& 5\arcsec$\times$5 \arcsec	& 0.55$\pm$0.08	& --27$\pm$2\\
V1515 Cyg	& $J$	& Oct 6, 2014 	& 20\arcsec$\times$10\arcsec	& 1.5$\pm$0.1		& --3$\pm$1\\
			& $H$	& Oct 5, 2014 	& 20\arcsec$\times$10\arcsec	& 1.16$\pm$0.02	& --13$\pm$1\\
\tableline \tableline
\end{tabular}
\end{table*}


    Data reduction implemented the standard procedures for ADI+PDI \citep{Hinkley09}, which includes flat fielding, removal of hot and bad pixels, distortion correction, image re-alignment, stacking multiple frames and calculating the Stokes parameters $I$, $Q$, and $U$. We then calculated the polarization intensity (PI=$\sqrt[]{Q^2 + U^2}$) and angle of polarization ($\theta_{p} = 0.5 \times \mathrm{arctan}(U/Q)$). The instrumental polarization of HiCIAO at the Nasmyth platform was corrected following \citet{Joos08} with errors of $<$0.1 \%.
    
    As described in Section 1, our target emission is polarized scattered light in a dusty circumstellar disk and/or an envelope illuminated by the central source, i.e., an unresolved inner disk associated with the protostar \citep[see][for reviews]{Hartmann96,Audard14}. The observed PI intensity distribution is therefore proportional to the $I$ flux from the central source. We measured the $I$ flux of the bright central unresolved source using short exposures, scaled to the long exposures, then normalized the PI intensity distribution in the science images by this $I$ flux  from the central point source (hereafter $I_*$). This will enable us to discuss the observed scattered flux and the intensity distribution independent of the brightness of the central illuminating source.
    
\begin{figure*}
\plotone{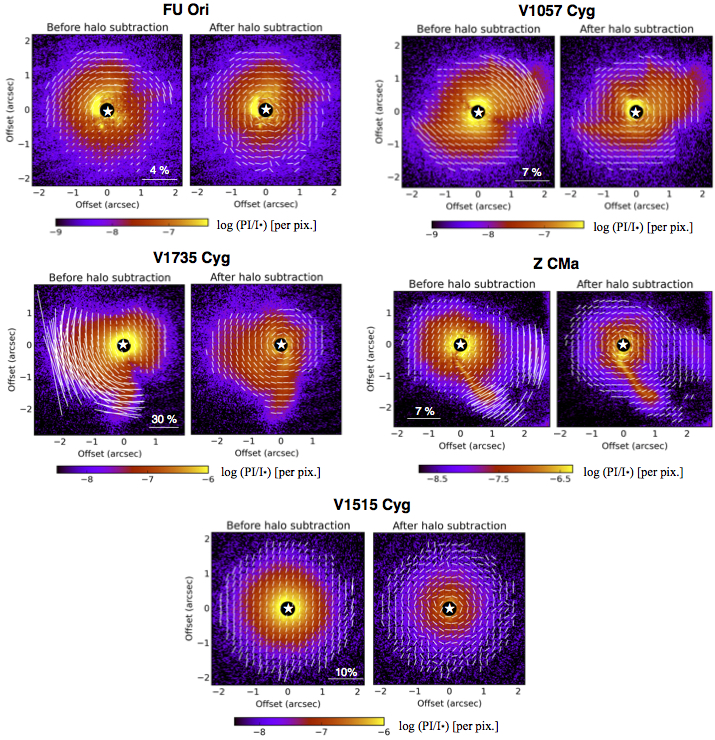}
\caption{The distribution of polarized intensity (PI intensity, color) and polarization vectors in $H$-band (1.65 \micron) before and after removing the polarized halo. 
For each object, the images before and after the halo correction are shown in the left and right side, respectively.
These images are convolved using a Gaussian with a FWHM=0\farcs02 to increase the signal-to-noise.
We set a software aperture for the coronagraphic mask of 0\farcs4 in diameter, slightly larger than that in the optics (0\farcs3 in diameter) to show the PI intensity distribution only where the measurements are reliable.
The PI intensity measured in each detector pixel is normalized to the stellar $I$ flux ($I_*$), and shown in a logarithmic scale. \\
The length of the polarization vectors in the left image represents the degree of polarization.
The PI intensity distribution in V1515 Cyg is centrosymmetric, and the polarization vectors show nearly the same orientation and degree of polarization over the entire region, indicating that the flux is dominated by the halo of the weakly-polarized central source. We use a constant length for the polarization vectors in the right image to clearly demonstrate the centrosymmetric pattern and therefore the halo correction goes well. 
The north is up for all the images.
\label{fig:halo_correction}}
\end{figure*}

    The bright central unresolved source is  very weakly polarized due to scattering in this region and/or dichroic absorption in the foreground circumstellar envelope or a parent molecular cloud. A halo of this unresolved source, caused by imperfect adaptive optics performance, extends over the disk and/or the envelope region. This contamination is observed as an alignment of polarization vectors toward a single direction, in particular close to the central soruce (Figure \ref{fig:halo_correction}). As for Paper I and some other work \citep[e.g.,][]{Hashimoto12,Follette13}, we subtracted this halo for FU Ori, V1057 Cyg, V1735 Cyg and Z CMa as follows. We assume the observed polarization is due to (1) an intrinsic centrosymmetrical circular pattern of polarization around the central object, as expected for single scattering from circumstellar dust grains; and (2) a halo or an outskirt of the point-spread function with a single direction of polarization vectors and a single degree of polarization. We then used a least-square method to optimize $q_*=Q_*/I_*$ and $u_*=U_*/I_*$, where $I_*$, $Q_*$ and $U_*$ are Stokes parameters of the central source. See Appendix \ref{app:optimization} for details of the optimization and validity of the assumptions. 
    
    Figure \ref{fig:halo_correction} shows the PI image of these objects for $H$-band before and after this correction. The length of the polarization vectors in the former image represents the degree of polarization, while we use a constant length for the polarization vectors in the right image to clearly demonstrate a centrosymmetric pattern and therefore that the halo correction went well.
    In Table \ref{tbl:observations} we also show the degree and the position angle (P.A.) of the polarization of each source and band derived from $q_*$ and $u_*$ as described above. In Appendix \ref{app:origin_halo_polarization} we briefly discuss the implications of these polarizations.
    

 	The PI intensity distribution observed in V1515 Cyg is centrosymmetric in the left panel in Figure \ref{fig:halo_correction} (shown in color), while the polarization vectors in the same panel show almost the same degree and angle of polarization over the entire region. These indicate that the PI flux is dominated by the polarized halo described above before its subtraction. The PI intensity distribution after the halo subtraction looks almost the same as before the halo subtraction but with a significantly lower intensity level. While this may be real signals due to scattering on a face-on disk or an envelope, we cannot exclude the possibility that it is an artifact of imperfect subtraction. We therefore regard the PI emission after the halo subtraction as an upper limit of the observations.


\section{Results}

Figures \ref{fig:fuori}--\ref{fig:zcma} shows the PI images of FU Ori, V1057 Cyg, V1735 Cyg and Z CMa, respectively. While the PI intensities are shown with a unit of per pixel for PI/$I_*$ in Figure \ref{fig:halo_correction}, as in \citet{Takami13,Takami14,deLeon15}, we use the unit of AU$^{-2}$ for Figures \ref{fig:fuori}--\ref{fig:zcma} as in \citet{Ohta16} for the following reason. While the PI flux per pixel is independent of the target distance, $I_*$ is inversely proportional to the square of the target distance. Therefore, PI/$I_*$ per pixel is proportional to the square of the target distance. We can cancel out such a dependency on distance if we use the unit of AU$^{-2}$. This will allow us to compare the observed intensity distribution and flux between different objects (Section 3.5). 

The observed extended emission is due to scattered light in a dusty circumstellar disk and/or an envelope illuminated by the central source. The emission is therefore fainter in the outer regions due to fainter illumination from the star. To clearly show faint extended structure in the outer regions, we multiplied the PI intensity at each position by $R^2$, where $R$ is the projected distance from the center of the coronagraphic mask, which approximately corresponds to the position of the central source, in arcsec. This image is shown in the right side of each figure.

The PI and PI$\times$$R^2$ images are convolved with Gaussians of FWHM=0\farcs02 and 0\farcs15, respectively, for better signal-to-noise. A modest FWHM (0\farcs02) is used for the PI images so as not to degrade the angular resolution of the observations ($\sim$0\farcs08), while a larger FWHM (0\farcs15) is used for PI$\times$$R^2$ to show faint emission in the outer region with a better signal-to-noise.

\begin{figure*}
\epsscale{0.7}
\plotone{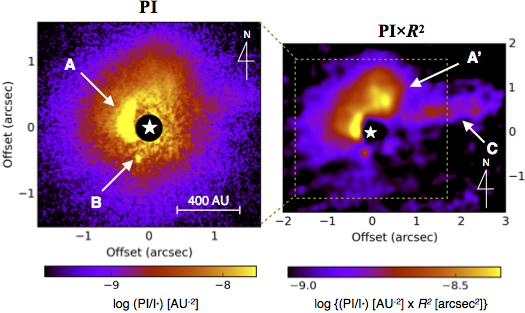}
\caption{The halo-corrected PI images of FU Ori ($H$-band, 1.65 \micron). The intensity at each position is normalized to the stellar $I$ flux ($I_*$). As for Figure \ref{fig:halo_correction}, we set a software aperture for the coronagraphic mask of 0\farcs4 in diameter, slightly larger than that in the optics (0\farcs3 in diameter) to show the PI intensity distribution only where the measurements are reliable.
For the right image we multiplied all pixels in the halo-corrected PI image by $R^2$, where $R$ is the projected distance from the center of the mask in arcsec.
The PI intensity is shown in a logarithmic scale in the unit of AU$^{-2}$. The features discussed in the text are marked as *A*, *A'*, *B*, and *C*.
\label{fig:fuori}}
\end{figure*}

    
\subsection{FU Ori (Figure \ref{fig:fuori})}

In the PI image we clearly find the following features reported in Paper I. These are: (A) an arc in the east like a spiral arm; and (B) an unresolved source corresponding to the companion FU Ori S. The component (B) is probably associated with compact thermal dust emission observed with millimeter to centimeter interferometry by \citet{Hales15,Liu17}. \citet{Liu17} show that the 1-cm dust continuum emission is dominated by an unresolved source at a 0\farcs08$\times$0\farcs07 resolution. This dust component may be due to a compact circumstellar disk associated with FU Ori S.

In the PI$\times$$R^2$ image we newly identify an emission component to the north (*A'* in the figure), which may be an outer extension of the arc *A*. The image also shows a tail at $\sim$3\arcsec~ ($\sim$1300 AU) to the west (*C* in the figure), which is seen only marginally in Paper I. The direction of the tail *C* is quite different from that of an outflow cavity, which is seen to the northeast in optical scatted light at an arcminute scale \citep{Hartmann96}. The full width zero intensity of the tail along the vertical axis may be 1\arcsec--1\farcs5~(400-600 AU), but a higher signal-to-noise is required for an accurate measurement.


\begin{figure*}
\epsscale{1.1}
\plotone{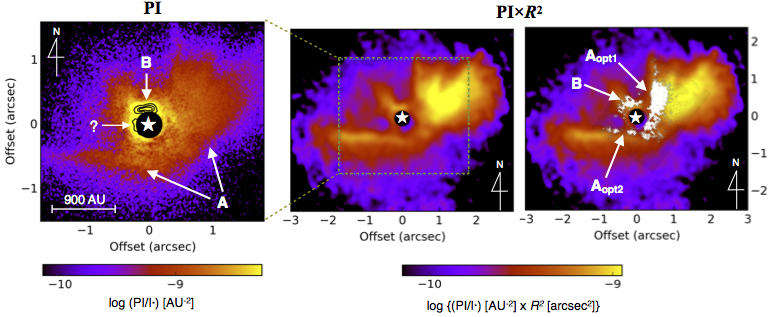}
\caption{Same as Figure \ref{fig:fuori} but for V1057 Cyg observed at $H$-band (1.65 \micron). In the PI image in the left, we show the black contours near the central source with arbitrary scales to clarify the small arc-like structure marked as *B*. We show PI$\times$$R^2$ images in the middle and the right, and the white regions in the right image show the distribution of optical emission observed by \citet{Herbig03} using the {\it Hubble Space Telescope}. 
\label{fig:v1057}}
\end{figure*}

\subsection{V1057 Cyg (Figure \ref{fig:v1057})}

As shown in Paper I, the observed intensity distribution consists of (A) spiky features extending to position angles between --60\arcdeg to 90\arcdeg; and (B) a small arc-like structure just to the north of the star. The PI$\times$$R^2$ image confirms an angular scale of the spiky features of $\sim$1\arcsec~from the star. The PI image also shows a marginal extension from the central source (the software mask) to the east, as clarified using contours in the PI image and marked with ``?". Observations with a better inner working angle are necessary to confirm if this feature is real, or an artifact due to imperfect adaptive optics correction.

\citet{Herbig03} observed the optical scattering nebulosity using the {\it Hubble Space Telescope}, and we superimpose their image after subtracting the stellar halo (the bottom-right image of their Figure 15) on the right image panel in Figure \ref{fig:v1057}. The optical emission is associated with *B*, and the north and the east inner edges of the spiky features *A* (*A$_{\mathrm opt1}$*, *A$_{\mathrm opt2}$*).
The optical image also shows an arc-like nebulosity at 2\farcs5--6\arcsec away from the star toward the south to southwest, but our image does not show emission associated with this feature.



\begin{figure*}
\epsscale{0.7}
\plotone{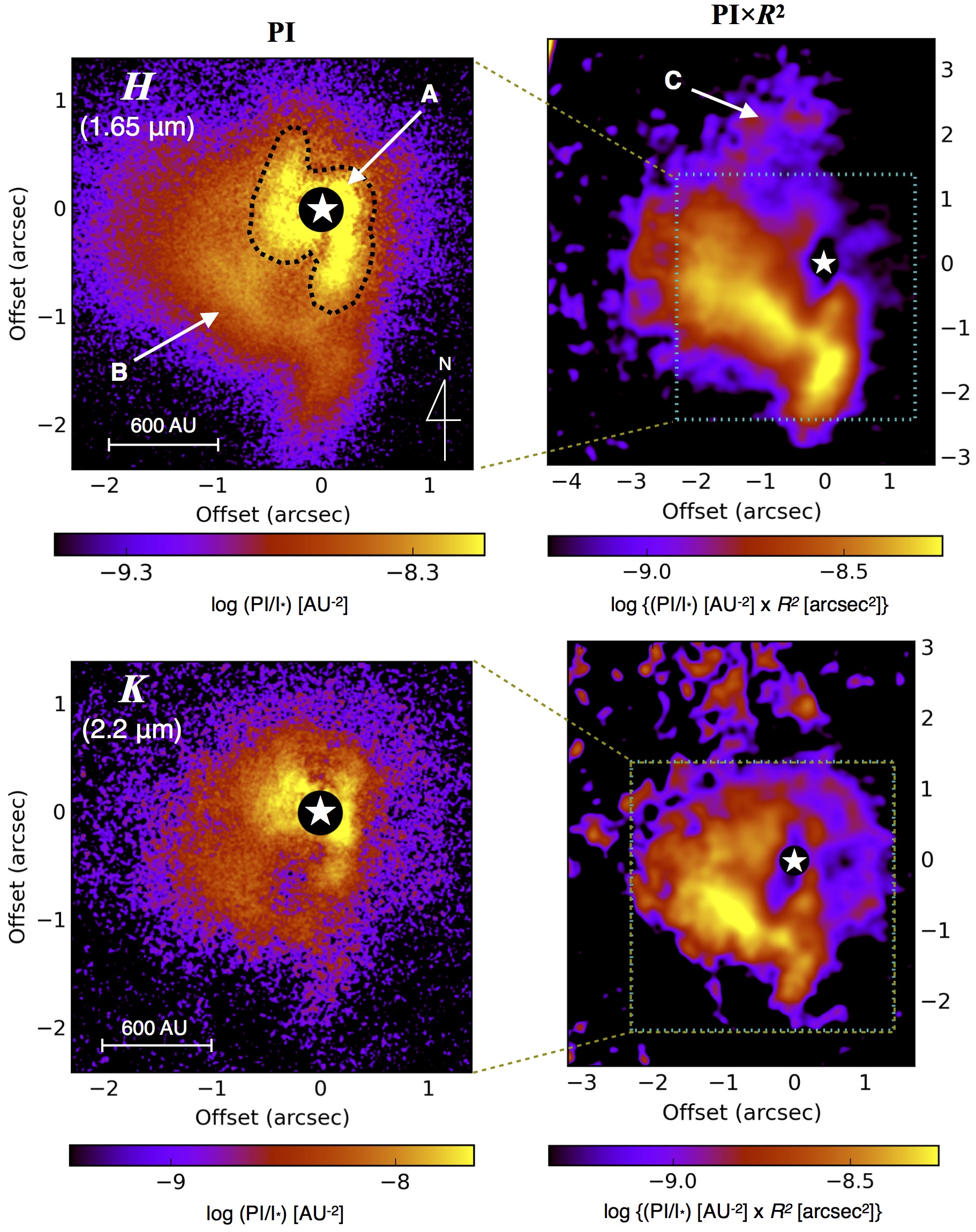}
\caption{Same as Figure \ref{fig:fuori} but for V1735 Cyg observed at $H$-band (upper, 1.65 \micron) and $K$-band (lower, 2.2 \micron).
The dotted curves in the top left image show the possible spiral structure discussed in the text.
\label{fig:v1735}}
\end{figure*}

\subsection{V1735 Cyg (Figure \ref{fig:v1735})}
The PI and PI$\times$$R^2$ images in the $H$- and $K$-bands (1.65 and 2.2 \micron, respectively) show complex fragmented structures. The PI image in the $H$-band (top-left) shows structures like spiral arms in the bright region within 0\farcs7 of the central source (*A* in Figure \ref{fig:v1735}). These structures are marginally seen in $K$-band (bottom-left of Figure \ref{fig:v1735}) as well. The PI and PI$\times$$R^2$ images also show faint extended emission in the east-to-south direction over a 2-arcsec scale from the central source (*B* in Figure \ref{fig:v1735}). In addition, the PI$\times$$R^2$ images show marginal extended emission from the star to the north at a $\sim$3-arcsec scale (*C* in Figure \ref{fig:v1735}).

Paper I suggested the presence of bright ``arms" extending to the northeast and southwest from the star. In our updated images, the shapes of these structures change considerably. This is due the fact that the subtraction of polarized halo was not perfect in Paper I (Appendix \ref{app:optimization}).


\begin{figure*}
\epsscale{0.7}
\plotone{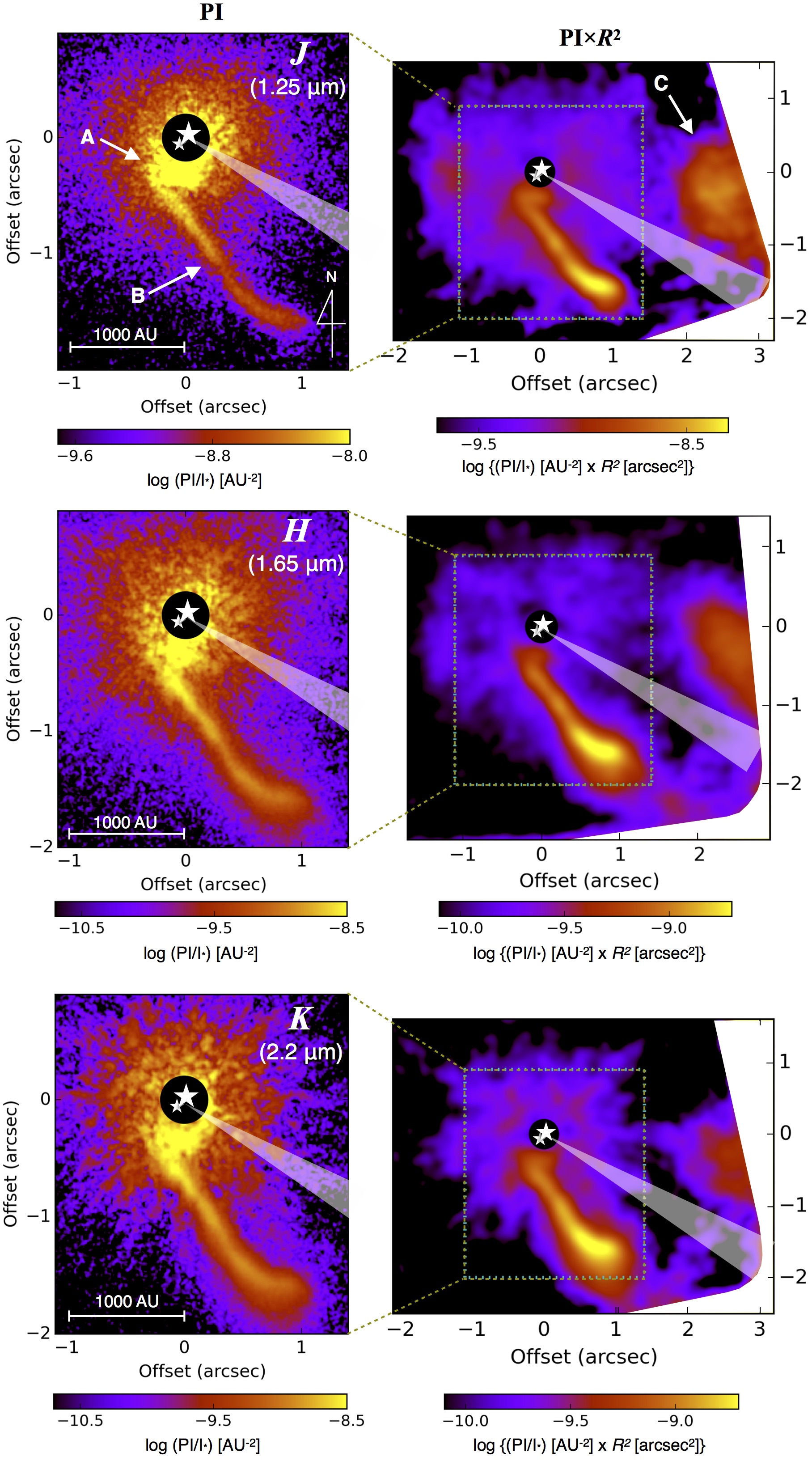}
\caption{Same as Figure \ref{fig:fuori} but for Z CMa observed at $J$- (1.2 
\micron, upper), $H$- (1.65 \micron) and $K$- bands (2.2 \micron, lower). Two binary components are shown in the software mask based on our measurement using short exposures ($d$=0\farcs11, P.A.=136\arcdeg). The location of the individual binary components in the mask is not accurately known due to the opaque coronagraphic mask. The half-transparent white triangles show the direction of two jets associated with the two binary components \citep[P.A.=$235\arcdeg$ and  $245\arcdeg$;][]{Whelan10}. The features discussed in the text are marked as *A*, *B*, and *C* in the top images. 
\label{fig:zcma}}
\end{figure*}

\subsection{Z CMa (Figure \ref{fig:zcma})}

The PI images obtained in the $J$-, $H$-, and $K$-bands (1.25, 1.65 and 2.2 \micron, respectively) show (A) a bright emission component extending to the south to $\sim$0\farcs4 from the star; and (B) an elongated structure in the south at a $\sim$2\arcsec~scale. These have been shown through previous near-infrared imaging and imaging polarimetry \citep[][; Paper I]{Millan-Gabet02,Canovas15}. As reported in previous publications, its position angle from the star is different from those of known jets observed by \citet{Whelan10} by 20\arcdeg--30\arcdeg. The PI$\times$$R^2$ images at three wavelengths also show faint extended emission to the west (*C*), offset from the jet direction by 20\arcdeg--30\arcdeg~but in an opposite side from *B*.


\subsection{Comparison of PI Intensities And Fluxes Between Objects}

\begin{figure*}
\epsscale{1.0}
\plotone{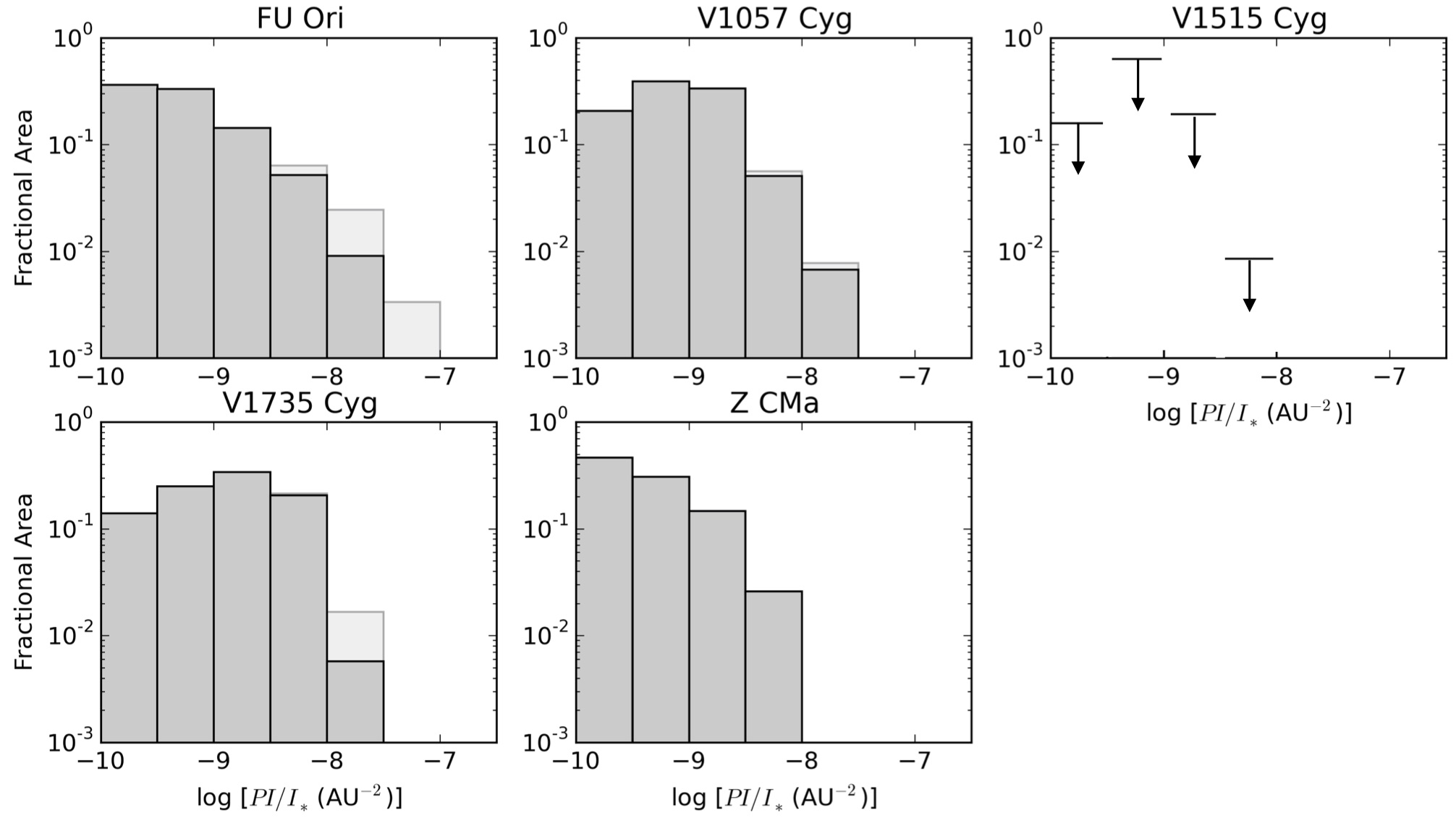}
\caption{The PI intensity distribution of the four objects in the $H$-band as histograms. The fractional areas for the individual bins were measured within 1000 AU of the central object. The central region with a radius of 200 AU is also excluded for the solid histograms. The faint histograms show the PI intensity distribution within 200 AU but outside the software mask ($r$=0\farcs2) used for Figures \ref{fig:halo_correction}-\ref{fig:zcma}. See Table \ref{tbl:PI_and_dust_masses} for the radii corresponding to $r$=0\farcs2 for the individual objects. The upper limit of the intensity distribution is provided for V1515 Cyg (see Section 2). For V1515 Cyg and Z CMa, the results for the regions of 200$<r<$1000 AU  and 0\farcs2$<r<$1000 AU are identical (as 0\farcs2 corresponds to 200 AU for these objects), therefore only the latter are shown in the figure. .
\label{fig:intensity_histograms}}
\end{figure*}

Figure \ref{fig:intensity_histograms} shows the PI intensity distributions of the above four objects in the $H$-band as histograms. We measure the fractional areas for the individual bins of the PI/$I_*$ intensities integrated over 200$<$$r$$<$1000 AU from the central object. Contributions at $r$$<$200 AU but outside the 0\farcs4-aperture software mask (cf. Section 2 and Figures \ref{fig:halo_correction}-\ref{fig:zcma}) are also shown as faint histograms to investigate how the innermost regions affect our estimate of the dust masses below. We also measure the total PI/$I_*$ fluxes integrated over these regions and tabulate it in Table \ref{tbl:PI_and_dust_masses}.

In Figure \ref{fig:intensity_histograms} and Table \ref{tbl:PI_and_dust_masses}, FU Ori, V1057 Cyg and V1735 Cyg show similar PI intensity distributions and fluxes, respectively. 
The spatially integrated PI/$I_*$ flux for V1735 Cyg is larger than FU Ori and V1057 Cyg by a factor of 1.4-2.
Those for Z CMa are lower than the other three objects, by a factor of 1.4-3 for the PI flux integrated over 200$<$$r$$<$1000 AU.


\begin{table*}
\caption{Spatially Integrated PI Fluxes and Dust Masses \label{tbl:PI_and_dust_masses}}
{\footnotesize
\begin{tabular}{lccccccc}
\tableline\tableline
Object
& \multicolumn{5}{c}{This Work}
& \multicolumn{2}{c}{Millimeter Observations}
\\

& $r_0$\tablenotemark{a}
& \multicolumn{2}{c}{$r_0$$<$$r$$<$1000 AU}
& \multicolumn{2}{c}{200 AU$<$$r$$<$1000 AU}
& \multicolumn{2}{c}{of Dust Mass ($M_\sun$)}
\\

& (AU)
& PI/$I_*$
& $M_{\mathrm dust}$ ($M_\odot$)\tablenotemark{b} 
& PI/$I_*$	& $M_{\mathrm dust}$ ($M_\odot$)\tablenotemark{b}
& \citet{Liu18}\tablenotemark{c}
& \citet{Sandell01}\tablenotemark{d}
\\

\tableline
FU Ori 		& 84	
& 4.3$\times$$10^{-3}$	& 1.4$\times$$10^{-6}$	& 2.7$\times$$10^{-3}$	& 1.3$\times$$10^{-6}$
& (2.9$\pm$0.4)$\times$$10^{-4}$	&(1.9$\pm$0.1)$\times$$10^{-4}$	\\

V1057 Cyg 	& 180
& 3.6$\times$$10^{-3}$	& 2.3$\times$$10^{-6}$ 	& 3.5$\times$$10^{-3}$	& 2.3$\times$$10^{-6}$
& (1.4$\pm$0.2)$\times$$10^{-4}$	& (1.0$\pm$0.1)$\times$$10^{-3}$ 	 	\\

V1735 Cyg 	& 120
& 6.5$\times$$10^{-3}$	& 4.2$\times$$10^{-6}$	& 5.7$\times$$10^{-3}$	& 4.2$\times$$10^{-6}$
&$<$2.8$\times$$10^{-4}$	& (4.2$\pm$0.6)$\times$$10^{-3}$ 	 	\\

Z CMa 		& 200
& 1.9$\times$$10^{-3}$	& 1.2$\times$$10^{-6}$	& 1.9$\times$$10^{-3}$	& 1.2$\times$$10^{-6}$
& (2.4$\pm$0.5)$\times$$10^{-3}$	& (1.4$\pm$0.1)$\times$$10^{-2}$ 	 	\\

V1515 Cyg 	& 200	
& $<$2.2$\times$$10^{-3}$ & $<$1.7$\times$$10^{-6}$  & $<$2.2$\times$$10^{-3}$ & $<$1.7$\times$$10^{-6}$ 
&(5$\pm$1)$\times$$10^{-4}$	& (1.3$\pm$0.1)$\times$$10^{-3}$ 	\\
\tableline \tableline
\end{tabular}
}
\tablenotetext{a}{Radius corresponding to that of the software mask at the central point source.}
\tablenotetext{b}{We do not expect accuracy of the tabulated values due to the uncertainty of the dust grain distribution along the line of sight (see text). The one digit decimal point is shown to demonstrate that the inner mask size would not significantly affect the estimate of the inferred dust mass.}
\tablenotetext{c}{Based on the 1.3-mm flux from the unresolved or marginally resolved source at a 1\arcsec-1\farcs5 resolution,  assuming an isothermal temperature of 20 K. The tabulated uncertainties are based on measurements of the millimeter fluxes and distances. The actual uncertainties may be larger due to uncertainty in the dust properties used to convert the millimeter fluxes to dust masses.}
\tablenotetext{d}{Based on the 0.5--1.3 mm fluxes from the central ``compact" source at a 8\arcsec-21\arcsec resolution, assuming an isothermal temperature of 50 K. We have scaled the values tabulated in the literature to the adopted distances listed in Table \ref{tbl:targets}. As the authors did not discuss the uncertainties in the dust masses, we include only those for the distance measurements. The actual uncertainties may be larger due to uncertainty in the dust properties used to convert the millimeter fluxes to dust masses.}
\end{table*}

We derive the dust mass inferred from the PI intensity distribution using the following equation \citep{Takami13}:
\begin{equation}
M_{\mathrm dust} = \int m({\bf r}) ~ d{\bf r} 
= \int \frac{\mathrm{PI} ({\bf r})}{I_*} \frac{r^2}{\kappa_{\mathrm ext}} \left( \frac{PI}{I_0} \right)^{-1} d{\bf r}, \label{eqn:dust_mass}
\end{equation}
where $m({\bf r})$ and PI are the dust mass and the observed intensity at each position, respectively;
$r$ is the distance  to the illuminating source; $\kappa_{\mathrm ext}$ is the extinction cross section; and (PI/$I_0$) is the fraction of the PI flux normalized to the incident flux on the dust grains. We use this equation with a standard interstellar dust model \citep[see Table \ref{tbl:dust} for summary; see also][]{Takami13}. For $r$ we use the projected distance as the actual geometry along the line of sight is not known.
Because the projected distance is a lower limit of $r$, it may cause an uncertainty of the derived dust mass by a factor of 2-3.

\begin{table*}
\caption{Assumed Dust Properties \label{tbl:dust}}
\begin{tabular}{ll}
\tableline\tableline
Parameter 	& Value \\
\tableline
Composition 					&	Silicate \& graphite\tablenotemark{a}\\
Mass Fraction \& Size Distribution 	& \citet{Kim94} \\
Mass Density					&	3.3/2.26 g cm$^{-3}$ for silicate/graphite\\
$\kappa_{\mathrm ext}$  ($H$-band)	&	5.3$\times$$10^3$ cm$^2$ g$^{-1}$ \tablenotemark{b}\\
PI/$I_0$  ($H$-band)				&	0.01\tablenotemark{b}\\
\tableline \tableline
\end{tabular}
\tablenotetext{a}{Different authors use different types of carbon dust, either graphite and amorphous carbon. While graphite has been extensively used, far-infrared SEDs of young stellar objects and evolved stars suggest the absence of graphite and the presence of amorphous carbon in circumstellar dust \citep[, and references therein]{Jager98}.  We use graphite here as \citet{Kim94} measured the size distribution of dust grains assuming graphite for the carbon dust. }
\tablenotetext{b}{\citet{Takami13}. The PI/$I_0$ is a typical value shown in Figure 5 of the paper.}
\end{table*}

The estimated dust masses are tabulated in Table \ref{tbl:PI_and_dust_masses}. We find a trend similar to the spatially integrated PI fluxes. The difference between FU Ori and V1057 Cyg is only a factor of 1.6--1.8. The dust mass estimated for V1735 Cyg is 2--3 times larger than those of FU Ori and V1057 Cyg, while that of Z CMa is smaller than these two objects by a factor of 1.1--3. Although the region closest to the central source is not included in our measurement and estimate, it may not have much effect, in particular for our estimate of the dust mass. For FU Ori, the PI flux integrated over 0\farcs2$<$$r$$<$200 AU (84$<$$r$$<$200 AU) is responsible for $\sim$40 \% of that integrated over 0\farcs2$<$$r$$<$1000 AU (84$<$$r$$<$1000 AU), while the dust mass integrated over the former region is only 6 \% of the latter region. The discrepancies of the PI fluxes integrated over the two regions are comparable or smaller for the other objects.

Does the above result imply that V1735 Cyg hosts a relatively dust-rich environment, while Z CMa hosts a relatively dust-poor environment? We cannot conclude this based on our near-IR observations, as the circumstellar disk plus the envelope can be optically very thick at these wavelengths \citep[e.g.,][]{Whitney03b,Kospal08,Takami13,Takami14}. In this context, the above dust masses is probably a small fraction of the entire dust mass illuminated at the surface of the circumstellar disk or the envelope. The circumstellar disks and the envelopes are optically much thinner at millimeter wavelengths, therefore we also list in Table \ref{tbl:PI_and_dust_masses} the dust masses based on \citet{Liu18,Sandell01} at millimeter wavelengths for comparisons. \citet{Liu18} performed interferometric observations of the 1.3-mm continuum toward 29 active YSOs including our targets. In their images the millimeter emission towards our targets is unresolved or only marginally resolved at a 1\arcsec--1\farcs5 resolution. This emission component is very likely to be associated with a (inner) disk region. We use the same formula as \citet{Liu18} ($M_{\mathrm dust}$ [$M_\sun$]=1.13$\times$$10^{-5}$ $F_{1.3mm}$ at $d$=350 pc) to convert the millimeter flux to the dust mass. \citet{Sandell01} performed single-dish observations at 0.5--1.3 mm with angular resolutions of 8\arcsec-21\arcsec, and measured fluxes at the ``central compact sources'' associated with 17 active YSOs including our targets.

The dust masses derived by \citet{Sandell01} are comparable to those based on \citet{Liu18} for FU Ori, V1057 Cyg, and V1515 Cyg, while the former are significantly larger than the latter for V1735 Cyg and Z CMa, i.e., by a factor of 5 or larger, due to contributions from an extended circumstellar envelope. This trend agrees with the fact that V1735 Cyg and Z CMa show silicate absorption at mid-IR wavelengths, indicating that these objects are surrounded by massive cold dust \citep{Quanz07}.

In contrast to the near-IR observations, the dust masses inferred from millimeter observations do {\it not} clearly show the trend that V1735 Cyg and Z CMa host dust-rich and dust-poor environments, respectively. These dust masses for Z CMa are significantly larger than the others, i.e., by a factor of 3--80.
Furthermore, the observations by \citet{Liu18} show that the dust mass associated with the V1735 Cyg disk is comparable or less massive than all the other objects but FU Ori. We use these discrepancies between the near-IR and millimeter observations to discuss a possible unified scheme with our targets and many other normal YSOs in Section 5.





\section{Implications for Individual Structures}

The observed morphologies in scattered light of the above four FUors are remarkably different from normal Class I-II YSOs. Class I YSOs are heavily embedded in the near-infrared, and the scattered light morphology often represents the shape of an outflow cavity in the protostellar envelope \citep[e.g.,][]{Tamura91,Lucas96,Lucas98a,Padgett99}. As these YSOs evolve toward the Class II phase, the dusty circumstellar envelope with an outflow cavity dissipates, allowing the star and the circumstellar disk to be directly observed even at optical wavelengths \citep[e.g.,][]{Whitney03b}. As a result, the observed scattered light from the Class II YSOs is dominated by the surface of the circumstellar disk \citep[e.g.,][for review]{Espaillat14}. Compared with those Class I-II YSOs, the scattered light morphologies shown in Section 3 are significantly more complex.

Furthermore, the observed morphologies in scattered light are significantly different between the four FUors, suggestive of different origins. We attribute them to (1) a gravitationally unstable disk; (2) ejection of gas+dust clumps; and (3) a wind or a jet, as described below in detail.

\subsection{A Gravitationally Unstable Disk}

The gravitational instability and resultant fragmentation of a circumstellar disk has been discussed for decades as one of the possible mechanisms to trigger FUor outbursts \citep[][for a recent review]{Audard14}. Hydrodynamical simulations have shown detailed structures of such disks, and how episodic mass accretion would occur \citep[e.g., ][]{Vorobyov10,Vorobyov15b,Machida11,Zhu12_GI,Meyer17,Zhao18}. \citet{Dong16} extended such simulations as made by \citet{Vorobyov15b} and simulated the distribution of near-infrared scattered light associated with these disks. Some of the images simulated by \citet{Dong16} show features and intensity distributions similar to the following features and morphologies observed in scattered light associated with FU Ori and V1735 Cyg. These are: (1) a one-arm spiral-like structure associated with FU Ori(*A* and *A'* in Figure \ref{fig:fuori}), as also discussed in Paper I; (2) bright structures like spiral arms associated with V1735 Cyg (*A* in Figure \ref{fig:v1735}); and (3) the fragmented extended structures associated with V1735 Cyg (*B* in Figure \ref{fig:v1735}).

One may alternatively attribute the near-infrared PI emission associated with V1735 Cyg to a remnant envelope, due to a relatively large mass inferred from near-IR observations (Table \ref{tbl:PI_and_dust_masses}).
However, there is no clear theory to explain the observed structures without gravitational instabilities.
Therefore, we suggest that this emission component is kinematically associated with a gravitationally unstable disk. Alternatively, the excess dust emission could be explained if the disk were flared more strongly than the existing models, and therefore receiving and scattering more photons from the star, and re-radiating more far-IR emission as well.

\subsection{Ejection of Gas+dust Clumps}

As shown in Figure \ref{fig:fuori}, FU Ori is also associated with an intriguing tail (*C*). This may be due to the ejection of gas+dust clumps from a gravitationally unstable disk. Hydrodynamical simulations by \citet{Vorobyov16b} show that fragments in a gravitationally unstable disk interact, and some of them can be ejected from the gravitational potential of the star+disk system.

\begin{figure}
\epsscale{1.2}
\plotone{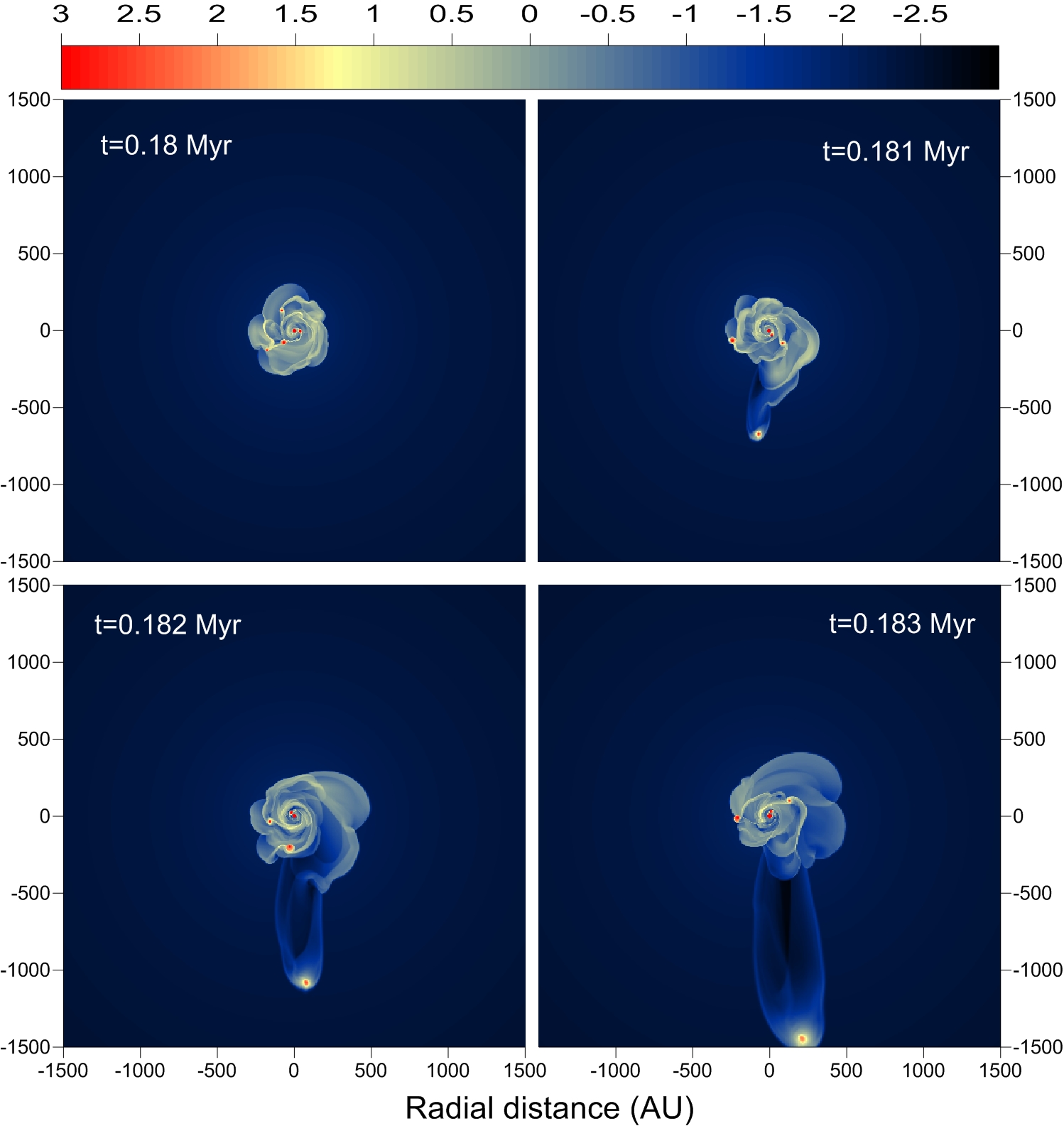}
\caption{Gas surface density maps (in log g cm$^{-2}$) showing the ejection of a dense gaseous fragment from the gravitationally unstable disk in numerical hydrodynamic simulations of \citet{Vorobyov16b}. The unstable disk hosts several fragments and the ejection occurs  through multi-body gravitational interaction between the fragments and the central star. The figure shows only a small fraction of the entire disk evolution and the time is counted from the formation of the disk.
\label{fig:Eduard1}}
\end{figure}

Figure \ref{fig:Eduard1} shows an example of such simulations executed by \citet{Vorobyov16b}, and extended by \citet{Vorobyov18} to include the evolving dust component. In the figure, the ejected clump is associated with a relatively straight tail, like tail *C* associated with FU Ori,  at a $\sim$1000 AU scale.
The mass of small dust responsible for scattered light in the near-IR is estimated to be $\sim$1.5$M_\earth$ for the tail of the ejected clump in Figure \ref{fig:Eduard1}. This is consistent with our observations of tail *C* associated with FU Ori. The dust mass inferred from the observed PI intensity distribution and Equation (\ref{eqn:dust_mass}) is $\sim$0.1 $M_\earth$ for the optically thin regime, and therefore $\gtrsim$0.1 $M_\earth$ for general case.


Alternatively, one might think that the observed tail *C* could be explained by a collimated jet, which is observed toward many YSOs in emission lines at UV to millimeter wavelengths \citep[e.g.,][for a recent review]{Frank14}. However, the direction of the tail is quite different from that expected from the geometry of the outflow cavity, or a possible Keplerian disk or envelope. \citet{Hartmann96} show an optical reflection nebula, analogous to an outflow cavity, extending to the northeast at an arcminute scale. Such a outflow cavity is well aligned with a jet for many YSOs \citep[e.g.,][for a review]{Reipurth01}.

Recently, \citet{Hales15} have marginally resolved kinematics in millimeter CO emission within 1 arcsec of the FU Ori binary system, showing that blueshifted and redshifted components lie across the binary orientation of P.A.$\sim$160$\arcdeg$.  These authors suggest the possibility that the CO emission they observed is associated with a Keplerian disk or an envelope. In this case one would expect the direction of the jet to be between north and northwest, or between south and southeast, as the rotational axis of a disk or an envelope often lies in the direction of a jet or an outflow \citep[e.g.,][for a review]{Frank14}. This direction is, again, quite different from that of tail *C*, which lies to the west.

An elongated structure similar to *C* toward FU Ori is also associated with Z CMa (*B* in Figure \ref{fig:zcma}), perhaps due to the same physical origin. 
The dust mass inferred from the observed PI intensity distribution and Equation (\ref{eqn:dust_mass}) is $\sim$0.7 $M_\earth$ for the optically thin regime, and therefore $\gtrsim$0.7 $M_\earth$ for general case, consistent with the modeled value described above.
The presence of a marginal curve at $>$1300 AU from the star agrees with hydrodynamical simulations by \citet{Vorobyov16b}, which show a relatively straight tail in the inner region (Figure \ref{fig:Eduard1}) but more complicated morphology in the outer region because of stretching caused by differential rotation.

The structure *C* associated with Z CMa was first discovered by \citet{Millan-Gabet02}, and the authors argued that the structure is associated with a wall of an outflow cavity. However, this explanation faces the following two difficulties. First, the observed morphology is quite different from scattered light in an outflow cavity observed toward many other YSOs. In general, the outflow cavity seen in scattered light is significantly more extended in the direction across the outflow axis \citep[e.g.,][]{Tamura91,Lucas96,Lucas98a,Padgett99}. The explanation of a cavity edge would also be excluded for the reason below. \citet{Nakajima95} conducted coronagraphic observations of Z CMa at optical wavelengths, and showed optical extended emission with a significantly wider opening angle (120\arcdeg-150\arcdeg) and a large angular scale ($\sim$40\arcsec). These structures can naturally be attributed to scattered light at a cavity edge, and it is unlikely that there is another thin cavity edge inside like the structure *B* associated with Z CMa.

Throughout, it is more likely that the tail-like structures associated with FU Ori (*C*) and Z CMa (*B*) result from the processes associated with a gravitationally unstable disk than a jet or an outflow cavity. It is noteworthy that hydrodynamical simulations by \citet{Vorobyov17} show that a stellar encounter with a gravitationally unstable disk also result in creation of similar tail structures. FU Ori and Z CMa are actually associated with a stellar component which is usually assumed to be a close binary companion \citep[e.g.,][see also Section 3]{Garcia99,Wang04,Reipurth04,Whelan10,Beck12_FUOri,Canovas15} rather than an external intruder. However, their binarity has not been confirmed based on kinematics or spectroscopy, implying that these ``companion" stars may still be external intruders. Furthermore, the presence of an intruder star has not been so far discovered toward the other objects. In these contexts, our observations are consistent with this scenario so far, and therefore this physical mechanism would remain as a possible explanation to explain the tail-like structures associated with FU Ori and Z CMa.

\subsection{A Wind/Jet}

The scattered light image obtained toward V1057 Cyg (Figure \ref{fig:v1057}) shows several spiky features extending between the east and the northwest directions. Such a morphology in scattered light is different from that of the other FUors we observed (Figures \ref{fig:fuori}, \ref{fig:v1735}, and \ref{fig:zcma}). This could be explained if a wind toward the east and the northwest were interacting with the circumstellar disk and envelope and blowing dust grains (as well as gas) away, allowing scattered light to be observed. 
Figure \ref{fig:v1057} shows that the spiky features in the near-IR are significantly more azimuthally extended than those in the optical, behind the inner edges traced in the optical emission. This is explained if the dust grains observed in the scattered light are associated with a wind cavity. A smaller cross section for the near-IR allows the emission to be observed in deeper, and therefore outer regions than at the optical wavelengths.

An energetic wind from V1057 Cyg has been observed over decades as blueshifted absorption in optical line profiles \citep[e.g., ][]{Herbig03,Herbig09}. We estimate a dynamical time scale of the observed spiky feature of $\sim$15 years from the observed spatial scale ($\sim$1500 AU), and a radial wind velocity of $\sim$400 km s$^{-1}$ \citep[e.g., ][]{Herbig03,Herbig09}. This time scale is much shorter than the presence of the energetic wind, which was continuously observed at least through 1981--2004 \citep[e.g.,][]{Bastian85,Herbig03,Herbig09}. Therefore, the structures clearly shown in Figure \ref{fig:v1057} would result from a relatively recent interaction between the wind and the surrounding dust, while dust grains interacting with a wind in the past would have been dispersed at further distances.

The scattered light distribution associated with V1057 Cyg is significantly different from that of many normal Class I YSOs, for which scattered emission tends to fill in the outflow cavity \citep[e.g.,][]{Tamura91,Lucas96,Lucas98a,Padgett99}. Such a peculiar intensity distribution for the V1057 Cyg wind may be due to a non-uniform distribution of the circumstellar disk (like a gravitationally fragmenting disk), which would be embedded in the envelope and therefore not clearly seen at near-IR wavelengths. Such a disk would result in an azimuthally non-uniform distribution of dust grains as a result of the wind-disk interaction. Otherwise, such structures in the inner disk region would partially block the light, resulting in non-uniform azimuthal illumination of the outer region.

One might think that a non-axisymmetric wind ejection could alternatively explain the observed intensity distribution, but we regard it unlikely for the reasons below. \citet{Herbig03,Powell12} showed periodic time variations of $\sim$15 days in the blueshifed absorption of optical lines (i.e., signature of a strong inner wind) toward FU Ori. This suggests that the wind originates from the very inner disk 
($\sim$20 $R_\sun$ from the star for FU Ori; see the references above)
Although these observations are not for V1057 Cyg, one would naturally expect a similar spatial scale and a similar rotational period for the wind launching region, and such a short rotational period would make the wind rather uniform in the azimuthal direction during the dynamical timescale of $\sim$10 years estimated above.

In addition to the scattered light toward V1057 Cyg, the diffuse component of emission *C* close to Z CMa (Figure \ref{fig:zcma}) is seen close to the jet direction, suggesting that the emission is associated with an outflow or a wind, or interaction of a outflow/wind/jet with the circumstellar envelope. \citet{Canovas12} also observed optical scattered light in this direction but in the outer region (3\arcsec-9\arcsec~from the star). A limited signal-to-noise hampers more detailed discussion for the origin of these dust components.

\section{A Possible Unified Scheme}

As described in Section 1, FUor outbursts have been discussed as a potential key mechanism for many (if not all) low-mass stars to accrete their final masses. If this is the case, we have to be able to explain the following observational trends. While FU Ori bursts may occur during the Class I phase, why do FUors we observed look very different from many normal Class I YSOs in the near-IR? Also, why do the FUors we observed look very different from each other in the near-IR? These can be explained if wind/jet ejections are episodic and/or time variable, as described below.

\begin{figure*}
\epsscale{1.}
\plotone{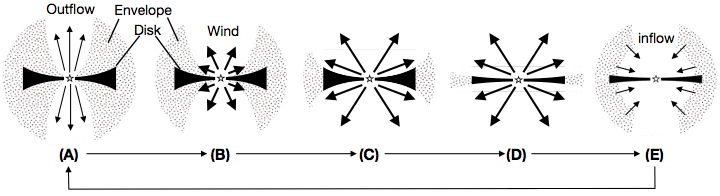}
\caption{The proposed sequence for the geometries of the disk and the envelope. See text for details.
\label{fig:sequence}}
\end{figure*}

Class I protostars are associated with a circumstellar disk, a circumstellar envelope and an outflow. These often exhibit a reflection nebula with a bipolar/monopolar morphology \citep[e.g.,][]{Tamura91,Lucas96,Lucas98a,Padgett99} associated with an outflow cavity. The circumstellar envelope exists below the wall of the outflow cavity. The disk cannot be seen in the near-IR because it is embedded in an optically very thick circumstellar envelope. See Figure \ref{fig:sequence} (A) for a schematic view of the disk, the envelope and the outflow cavity based on the above understanding \citep[e.g.,][for a review]{Stahler05}.


When an episodic energetic wind emerges, it would blow away the surface of the circumstellar disk and the inner region of the envelope at the spatial scale of our near-IR observations, i.e., $\sim$1000 AU (Figure \ref{fig:sequence}B). The disk surface would be seen when the circumstellar envelope has been blown away (FU Ori, V1735 Cyg; Figure \ref{fig:sequence}C). Scattered light from the disk becomes significantly fainter when this blowing process continues (Z CMa, V1515 Cyg; Figure \ref{fig:sequence}D). This process would not significantly affect the disk mass or the outer region ($\gg$1000 AU) of the envelope (Section 3.5).

The wind-ripping process also makes the tails, i.e., the trails of the clump ejections from gravitationally unstable disks, visible in the near-IR (FU Ori, Z CMa). In some cases the circumstellar dust being blown in the wind itself would be seen (V1057 Cyg, Z CMa). The non-uniform azimuthal distribution of the wind feature associated with V1057 Cyg may result from the wind-disk interaction of or uneven illumination associated with a gravitationally fragmenting disk.

When the wind stops, circumstellar material from the outer region would infall, quickly recovering the disk and the inner envelope (Figure \ref{fig:sequence}E). The free-fall time scale is only $\sim$5000 years at 1000 AU for a 1 $M_\sun$ central star. This is significantly shorter than the lifetime of the Class I phase \citep[$4-5 \times 10^5$ years, see][for a review]{Dunham14}. Therefore, the chance of observations of the scattered light distributions like those of FUors is very low during the Class I phase.

The repetition of the above processes gradually clears the outer envelope toward the Class II phase. This transition is observed in millimeter dust emission (Section 3.5) and millimeter CO emission at large scales \citep{Kospal17b}, and also via the mid-IR silicate feature with absorption and emission for early and late phases, respectively \citep{Quanz07}.


In summary, our results are consistent with the scenario that many low-mass YSOs experience FUor outbursts triggered by gravitational instability in disks in their Class I (and perhaps Class 0) evolutionary phases. This supports the conclusion of \citet{Dunham12}, who provided an explanation for the observed bolometric luminosities and temperatures with their gravitationally unstable disk models. 
Numerical simulations show that the fragmented structures of the disks are often present during these evolutionary stages \citep[e.g.,][]{Vorobyov05,Vorobyov10,Vorobyov15b,Boley10,Cha11,Tsukamoto13, Zhao18}, implying that many normal Class 0-I YSOs may also be associated with such gravitationally fragmented disks \citep[see e.g.,][for an example]{Tobin16}. However, disks associated with Class I protostars are embedded in the envelope and therefore not visible in the near-IR \citep[e.g.,][for a review]{Dunham14}. High resolution millimeter interferometry is free from the problems of scattering geometry, and suffers significantly less from the optical thickness issue. Survey observations of Class I YSOs with the Atacama Millimeter/Submilllimerter Array (ALMA) would therefore allow us to further investigate the above scenario.

A combination of high spatial+spectral resolution and a high sensitivity of ALMA would also be powerful for observing gas kinematics in the inner envelope of the FUor objects and FUor-like objects in near future. This will allow us to search for signatures of the blowing-away and recovering processes, and therefore test the unified scheme discussed above.


\section{Summary}

We present near-IR imaging polarimetry of five classical FU Ori-type objects (FU Ori, V1057 Cyg, V1515 Cyg, V1735 Cyg, Z CMa) with Subaru-HiCIAO. Scattered light associated with circumstellar dust was observed around at least four of them (FU Ori, V1057 Cyg, V1735 Cyg, Z CMa), and possibly around the remaining object (V1515 Cyg) as well. Their polarized intensity distribution shows a variety of morphologies with arms (FU Ori, V1735 Cyg), tails or streams (FU Ori, Z CMa), spikes (V1057 Cyg), and fragmented distributions (V1735 Cyg). These significantly differ from scattered light observed toward many other normal young stellar objects (Class I-II objects). The observed arms and fragmented structures are attributed to gravitationally unstable disks; the tails are attributed to the trails of clump ejections; spiky structures are attributed to dust blown by a wind or a jet. We have also detected unresolved polarized emission, which may be due to a compact circumstellar disk, associated with the companion of FU Ori (FU Ori S).

We can consistently explain our results with the scenario that their accretion outbursts are triggered by gravitationally unstable (and therefore often fragmenting) disks, and with the hypothesis that many low-mass young stellar objects experience such outbursts during the Class I phase. Class I protostars are associated with a disk, and also an envelope for most of this evolutionary phase. When an episodic energetic wind emerges, it would blow away the surface of the circumstellar disk and the inner envelope. The disk surface would be seen, as for FU Ori and V1735 Cyg, when the inner circumstellar envelope has been blown away. Scattered light in the disk and the envelope would become significantly fainter, as for Z CMa and V1515 Cyg, after this outflow process continues. In some cases the circumstellar dust being blown in the wind itself would be seen (V1057 Cyg, Z CMa). When the wind stops, circumstellar material from the outer region would infall, quickly recovering the disk and the inner envelope in a time scale of $\sim$5000 years. 




\acknowledgments

We are grateful to the anonymous referee for useful comments.
We thank the Subaru Telescope staff for their support.
MT is supported from Ministry of Science and Technology (MoST) of Taiwan (Grant
No. 103-2112-M-001-029; 104-2119-M-001-018; 105-2112-M-001-023; 106-2119-M-001-026-MY3). 
E. Vorobyov acknowledges support from the Russian Science Foundation grant 17-12-01168.
AK has received funding from the European Research Council (ERC) under the European Union's Horizon 2020 research and innovation programme under grant agreement No 716155 (SACCRED).
This research made use of the
Simbad database operated at CDS, Strasbourg, France, and the
NASA's Astrophysics Data System Abstract Service.

%

\vspace{5mm}
\facilities{Subaru (HiCIAO)}


\software{IRAF \citep{Tody86,Tody93}, PyRAF \citep{pyraf}, numpy \citep{numpy}, scipy \citep{scipy}
          }



\appendix

\section{Distance to Z CMa \label{app:zcma_distance}}
The Gaia Data Release 2 (DR2) measured the parallax of this star to be 4.299$\pm$0.891 milliarcsec, yielding a distance of 230$^{+60}_{-40}$ pc. This contrasts to the photometric distance measured for this association \citep[1000--1300 pc][]{Claria74,Humphreys78,Kaltcheva00}. We are concerned that the Gaia measurements in 2014--2016 might have suffered significantly from photometric variabilities of this binary system with a $d=0\farcs11$ separation \citep{Bonnefoy17}. We therefore assume $d$$=$1000 pc based on \citet{Kaltcheva00}.

The above Gaia distance would significantly change the total mass of the binary system and the total luminosity of the FUor component. 
\citet{Canovas15} measured a change in binary $\Delta$PA=10\fdg6$\pm$2\fdg1 and a separation $\Delta$a=0\farcs003$\pm$0\farcs003 in 14 years, performing comparison with the projected separation and the PA measured by \citet{Millan-Gabet02}. Assuming a face-on view of a circular orbit, these orbital parameters yield a total mass of the binary system of 4-10 $M_\sun$ with the adopted distance, and 0.03--0.2 $M_\sun$ with the Gaia distance, both including the uncertainty in the distance. The latter mass may make Z CMa a very peculiar FUor, however, there are few measurements of FUor masses \citep[cf.][for FU Ori]{Beck12_FUOri}, and the mass derived above depends on the assumed orbital eccentricity and inclination.  

\citet{Thiebaut95} measured the total luminosity of the FUor component to be 650 $\L_\sun$ adopting $d$$=$930 pc. Adopting $d$$=$230 pc, this would be 27 $L_\sun$, significantly lower than the other classical FUors tabulated in Table \ref{tbl:targets}) (100--500 $L_\sun$) but comparable or larger than HBC 722 and V2775 Ori, two of the FUors recently discovered \citep[0.7--28 $L_\sun$][]{Audard14}.

We regard $d$=1000 pc as more likely than the Gaia measurement based on the above discussion of the astrometric binary motion and the luminosity of the FUor component. However, we emphasize that neither gives crucial constraints on the distance to Z CMa.



\section{Removing Polarized Halo \label{app:optimization}} %

The observed spatial distribution of the Stokes parameters may be described as follows:
\begin{equation}
\left( \begin{array}{c} I_{obs}({\bf r_{\mathrm i}}) \\ Q_{obs}({\bf r_{\mathrm i}}) \\ U_{obs}({\bf r_{\mathrm i}}) \end{array} \right)
= \left( \begin{array}{c} I_s({\bf r_{\mathrm i}}) \\ Q_s({\bf r_{\mathrm i}}) \\ U_s({\bf r_{\mathrm i}}) \end{array} \right)
+ I_*({\bf r_{\mathrm i}}) \left( \begin{array}{c} 1 \\ q_* \\ u_* \end{array} \right)
\end{equation}
where $I_{obs}({\bf r_{\mathrm i}}),Q_{obs}({\bf r_{\mathrm i}}),U_{obs}({\bf r_{\mathrm i}})$ are the observed Stokes parameters; ${\bf r_{\mathrm i}}$ is the position; $I_s({\bf r_{\mathrm i}}),Q_s({\bf r_{\mathrm i}}),U_s({\bf r_{\mathrm i}})$ are the Stokes parameters of the scattered light from a disk or an envelope; $I_*({\bf r_{\mathrm i}})$ is the Stokes $I$ of the central unresolved source convolved by the point-spread function; $q_*$ and $u_*$ are the Stokes parameters of the central source normalized to Stokes $I$. From Equation (B1) we derive:

\begin{eqnarray}
Q_s({\bf r_{\mathrm i}}) & = & Q_{obs}({\bf r_{\mathrm i}}) - q I_*({\bf r_{\mathrm i}}), \\
U_s({\bf r_{\mathrm i}}) & = & U_{obs}({\bf r_{\mathrm i}}) - u I_*({\bf r_{\mathrm i}}),
\end{eqnarray}
The normalized polarization vector of the scattered light is described using the following equations:
\begin{equation}
{\bf v} ({\bf r_{\mathrm i}}) = \left( \begin{array}{c} \mathrm{cos} ~\theta({\bf r_{\mathrm i}}) \\ \mathrm{sin} ~\theta({\bf r_{\mathrm i}}) \end{array} \right),
\end{equation}
where the position angle $\theta ({\bf r_{\mathrm i}})$ is described by:
\begin{equation}
\theta ({\bf r_{\mathrm i}}) = \frac{1}{2} \mathrm{arctan} \frac{U_s({\bf r_{\mathrm i}})}{Q_s({\bf r_{\mathrm i}})}
                         =  \frac{1}{2} \mathrm{arctan} \frac{U_{obs}({\bf r_{\mathrm i}}) - u I_*({\bf r_{\mathrm i}})}{Q_{obs}({\bf r_{\mathrm i}}) - q I_*({\bf r_{\mathrm i}})},
\end{equation}
Here we assume that the scattered light in the disk and/or the envelope provides a centrosymmetric circular pattern about the central source, i.e.,
\begin{equation}
{\bf v} ({\bf r_{\mathrm i}}) \cdot {\bf r_{\mathrm i}} = 0.
\end{equation}
In nature, the polarization vectors show a centrosymmetric circular pattern about the illuminating source if the dust grains are spherical, and the effects of multiple scatterings are negligible \citep[e.g.,][]{vandeHulst81}. This is approximately consistent with polarization patterns in scattered light toward many young stellar objects and protoplanetary disks \citep[e.g.,][]{Lucas98a,Perrin09,Takami13}.

We derived $q_*$ and $u_*$ using Equations (B4)-(B6) and the least square method, minimizing $\sum_i [{\bf v}({\bf r_i}) \cdot {\bf r_i}]^2$ based on (B6). This optimization was made for each object using
the region within 1\arcsec~of the star, i.e., where the halo significantly affects the observed polarization.

\begin{figure}
\epsscale{0.5}
\plotone{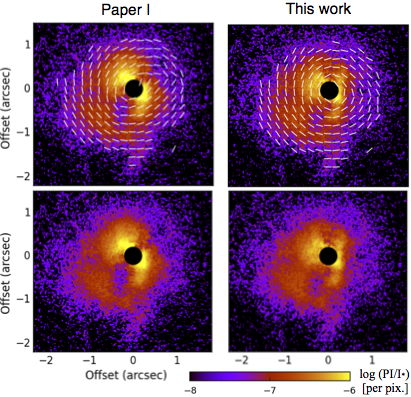}
\caption{PI intensity distribution and polarization vectors for Paper I and this work for the $K$-band observations of V1735 Cyg. The upper and lower figures are identical but the polarization vectors are removed for the latter to clearly show the PI intensity distribution.
\label{fig:vsPaperI_images}}
\end{figure}

\begin{figure}
\epsscale{0.5}
\plotone{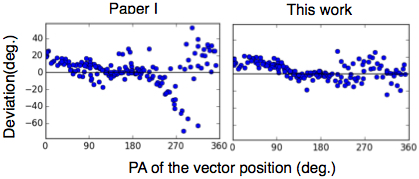}
\caption{Differences of polarization angles from the centrosymmetric pattern for the individual vectors in Figure \ref{fig:vsPaperI_images}. The horizontal axis is the position angle of the location of the individual vectors.
\label{fig:vsPaperI_plots}}
\end{figure}

With careful analysis, we were able to remove the polarized halo with an accuracy higher than Paper I for the $K$-band observations of V1735 Cyg. In Figure \ref{fig:vsPaperI_images} we show the PI intensity distribution and polarization vectors for Paper I and this work for this object and band. Although the vectors show a centrosymmetric pattern in both cases, that in Paper I is slightly aligned to the northeast-southwest direction to the northwest side of the star. This trend is clearly shown in Figure \ref{fig:vsPaperI_plots}, which shows the differences in polarization angles from the centrosymmetric pattern for individual vectors. The standard deviations from the centrosymmetric patterns are 17\fdg4 and 10\fdg7 for Paper I and this work, respectively.

In Paper I, the PI intensity distribution for for the $K$-band observations of V1735 Cyg shows bright emission extending to the northwest and southwest direction at a $\sim$0\farcs8 scale from the star. The shape of this structure changes considerably in the $K$-band image in the revised image. Paper I also shows the PI intensity distribution for FU Ori ($H$-band), V1057 Cyg ($H$-band), and Z CMa ($K$-band), and these now agree well with our revised halo correction.


\section{Origin of Weak Polarization of Central Sources \label{app:origin_halo_polarization}}

As described in Section 2, the bright central unresolved source is  very weakly polarized due to scattering in this region and/or dichroic absorption in the parent molecular clouds. The differences of the position angle and the polarization degree at multiple wavelengths would allow us to investigate its origin in more detail.

In V1735 Ori the position angle is the the same at $H$- and $K$-bands (18\arcdeg), while the polarization degree is larger in the former than the latter (1.7 \% and 1.0 \%, respectively; Table \ref{tbl:observations}). These trends could be explained by dichroic absorption in the parent molecular clouds, for which we expect a larger polarization at a shorter wavelength due to a large cross section for the dust grains. This explanation is corroborated by relatively large extinction ($A_V$=8--11; and therefore the column density of the gas+dust) toward this star, which would cause a larger polarization due to dichroic absorption than in the other stars \citep[$A_V$=2--4; ][; see Table \ref{tbl:targets} of this paper]{Audard14}.

For Z CMa, the position angle 
measured for $K$-band ($-27 \arcdeg \pm2 \arcdeg$) is slightly larger than those for $J$- and $H$-bands ($-21 \arcdeg \pm1 \arcdeg $ and $-23 \arcdeg \pm2 \arcdeg$, respectively). This trend could be explained with a combination of dust scattering near the central source and dichroic absorption in the parent molecular clouds, for which the position angles of polarization are different. We expect different position angles at different wavelengths as the wavelength dependence of polarization is different between these origins. This scenario would also explain the different polarization angles between $J-$ and $H-$ bands measured for V1515 Cyg ($-3 \arcdeg \pm1 \arcdeg $ and $-13 \arcdeg \pm1 \arcdeg$, respectively).

\end{document}